\documentclass[a4, 11pt]{article}
\usepackage[textwidth=16cm]{geometry}
\usepackage{amssymb, amsmath}
\usepackage{graphicx}
\usepackage{color,cite,authblk}

\newcommand{\fig}[1]{Fig.~\ref{#1}}
\newcommand{\eq}[1]{Eq.~(\ref{#1})}
\newcommand{\sect}[1]{Sec.~\ref{#1}}

\newcommand{\e}{\mbox{e}}
\newcommand{\dt}{\partial_{t}}
\newcommand{\dxs}{\partial_{x}}
\newcommand{\util}{\tilde{u}}
\newcommand{\ch}{\cosh}
\newcommand{\sh}{\sinh}

\begin{document}
\title{Trapping reaction in a symmetric double well potential}
\author[1]{Trilochan Bagarti\thanks{bagarti@hri.res.in}}
\author[2]{Kalyan Kundu}
\affil[1]{Harish-Chandra Research Institute\\ Chhatnag Road,Jhunsi, Allahabad-211019}
\affil[2]{Institute of Physics\\ Sachivalaya Marg, Bhubaneswar-751005}

\maketitle
\begin{abstract}
We study the trapping reaction-diffusion problem in a symmetric double well 
potential in one dimension with a static trap located at the middle of 
the central barrier of the double well. The effect of competition between 
the confinement and the trapping process on the time evolution of the survival 
probability is considered. The  solution for the survival 
probability of a particle is obtained by the method of Green's function. 
Furthermore, we study trapping in the presence of a growth term. We show 
that for a given growth rate there exist a threshold trapping rate beyond 
which the population can become extinct asymptotically. Numerical simulations 
for a symmetric quartic potential are  done  and  
results are discussed. This model can be applied to study the dynamics of 
a population in habitats with a localized predation.
\end{abstract}

\section{Introduction}
The trapping reaction diffusion model is one of the simplest models of 
diffusion limited reaction process. It has been used to describe a wide 
variety of phenomena such as the trapping of excitons in crystals, 
recombination of electron and hole, formation of soliton-antisoliton pair, 
reaction activated by catalysis\cite{hav}. The trapping reaction diffusion 
process consists of (i) free diffusion of a particle $A$ and (ii) the 
absorption of a particle whenever it encounters a trap $T$. The trapping 
reaction $A+T \rightarrow T$ shows anomalous kinetics that depends on the 
spatial dimension. For the simplest case of a single trap, the number of 
particles $n(t)$ decay as $t^{-1/2}$ in one dimensional space $d=1$ and 
$1/\log t$ when $d=2$\cite{redner1}.

The trapping problem with multiple traps randomly distributed in space has 
also been studied intensively in the past. In the asymptotic time limit it 
has been shown that, for static traps, $n(t)$ decays as a stretched exponential 
$\exp(-\alpha_d \rho^{2/(2+d)} t^{d/(2+d)})$, where $\alpha_d$ is a constant 
that depends on the spatial dimension $d$ and $\rho$ is the mean density of 
traps\cite{balagurov,donsker,gras}. The stretched exponential behavior 
appears due to large trap free regions where particles can spend exponentially 
long time before they get absorbed at a trap\cite{gras}. Trapping reaction 
can induce self-segregation of the reactant species\cite{jayan, htait},
 self-organization around traps\cite{wei89,tait,jayan1,thm}. In 
the presence of volume exclusion it has been found that, an additional 
term appears in the stretched exponent which effectively slightly increases
(decreases) the trapping  rate\cite{tab,tab2}. Similar anomalous behavior is 
observed in quantum transport in the presence of traps. However it has been 
found that the stretched exponential behavior in the quantum regime is slower 
than its diffusive counterpart\cite{krap,parris1,parris2,jayan2}. Random 
multiplication of particles in a diffusive medium which models chemical 
reactions, evolution of biological species, etc have been 
studied\cite{jayan3,valle,ebling}. For a population which undergoes 
multiplication and decay at random positions in space, it has been shown 
by using the knowledge of density of state for disordered system, that the 
asymptotically $n(t)$ is exponential times the survival probability for 
the single trap trapping problem.

Recently, trapping of a diffusing particle in a harmonic potential has been 
studied with the trap located at various position relative to the bottom of 
the potential and the initial position of the particle\cite{spend}. This model 
was motivated by problems in biophysics such as photosynthesis, DNA stretching 
with optical tweezers, etc.  Trapping problem with a potential can also be 
studied in the quantum regime. 

In this paper we study the trapping reaction-diffusion model with an external 
confining potential. We consider a symmetric double well potential and a trap 
at the origin. Our motivation in this work comes from possible applications 
in ecology. We shall briefly discuss the scenario in \sect{react.diff} where 
we describe the reaction diffusion model. In \sect{trap}, we study by using 
Green's function technique, the trapping of a single particle in a square 
double well potential. Trapping with growth has been studied in \sect{trap2} 
and the competition between decay and growth is discussed. Finally, 
numerical results for a general quartic double well potential has been 
discussed in \sect{num}.
\section{Reaction-diffusion equations}
\label{react.diff}
We consider particles diffusing in the presence of a external confining 
potential $V(x)$ with a trap located at the origin. We assume a symmetric 
double well potential with a maximum at the origin.  When a particle 
encounters the trap during its motion it gets absorbed at a rate $\kappa$. Let 
us denote by $u(x,t)$ the density of particles at position $x$ at time $t$. 
The reaction-diffusion equation can be written as 
\begin{align}
 \dt u = D\dxs^2 u - \dxs(\phi u)-\kappa \delta(x) u + F(u),
\label{1}
\end{align}
where $D$ is the diffusion coefficient of the particle, 
$\phi(x):= -\bar{\gamma}^{-1}\dxs V(x)$ with $-\dxs V(x)$ being the force on the 
diffusing particle due to the potential and $\bar{\gamma}$ is damping constant 
such that $D = k_BT/\bar{\gamma}$. 
The term $-\kappa \delta(x) u$ describes the trapping reaction 
at the origin and $F(u)$ describes the growth of the population. 
The boundary condition is $\lim_{|x|\rightarrow \infty} u(x,t)=0$ and the 
initial condition is $u(x,0) = f(x)$.

{\it Application to ecology: } This simple model can be applied to a two 
species predator-prey model with a localized predation. The external 
potential can be used to model the habitats. A typical 
example could be the 
predator-prey dynamics of herbivore and crocodile, fishing by humans etc. The 
predator in these cases are localized in a small region where as the prey 
can diffuse from one habitat to other. The knowledge of the predator 
localized at some region can build a fear in the prey. This will  create 
a barrier that gives rise to  repulsion away from the predation region. A part 
of the potential can also arise from basic problems of accessibility between 
two habitats. Two villages connected by bad roads or by turbulent rivers can
be an example. The tendency of the prey to stay in herds can be incorporated 
by an attraction towards the center of the herd. The size of the herd may 
depend on the size of the habitat. These two effects can naturally be 
incorporated by a double well potential in the reaction-diffusion equation.
\section{Trapping of a single particle}
\label{trap}
\begin{figure}
\centering
\includegraphics[width=0.6\textwidth]{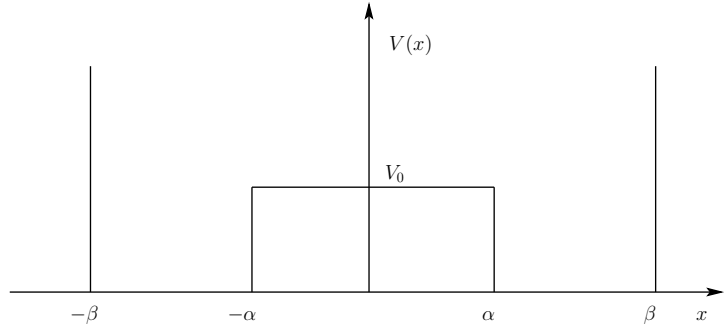}
\caption{Symmetric square double well potential.}
\label{fig.sqrdbwell}
\end{figure}

The trapping of a single particle in the presence of an external potential 
can be considered as an inhomogeneous pure death process, $F(u)=0$. Let us 
consider a square double well potential with a barrier height $V_0$ 
(see \fig{fig.sqrdbwell}) with a particle located initially in right well at $x=a$. 
Using the transformation $t \rightarrow Dt,~V\rightarrow$
$~V/k_BT,~\kappa \rightarrow \kappa/D,~$ in \eq{1} we obtain 
\begin{equation}
\dt u = \dxs^2 u - \dxs(\phi u) - \kappa \delta(x)u.
\label{1.1}
\end{equation}
Note that due to the said transformations, in \eq{1.1}, $V(x)$ is 
dimensionless and consequently $\phi = -\dxs V(x)$ has the dimension of 
length . Furthermore $t$ has the dimension of (length)$^{2}$ and $\kappa$ 
has the dimension of inverse length.
We gain note that the  Laplace Transform of \eq{1.1}  can be solved  
exactly by  the method of Green's function. Let $\util(x,s)$ be the 
Laplace transform of 
the the density $u(x,t)$ so that we have
\begin{equation}
\left[ s - \dxs^2 + \dxs \phi \right]\util(x,s) = u(x,0)-\kappa \delta(x)\util(x,s).
\label{2}
\end{equation}
The solution $\util(x,s)$ can be written as
\begin{equation} 
\util(x,s) = G(x|a) - \frac{\kappa G(x|0)G(0|a)}{1+\kappa G(0|0)},
\label{3}
\end{equation}
where $G = \left[s-\dxs^2-\dxs \phi \right]^{-1}$ is the Green's operator 
in the coordinate representation (see Appendix \ref{appndx}).
\subsection{Survival probability}
The survival probability $S(t)$ is defined as 
\begin{align}
S(t) &= \int u(x,t) dx.
\label{4}
\end{align}
From \eq{3} we obtain
\begin{equation}
\tilde{S}(s) = \frac{1}{q^2} - \frac{\kappa G(0|a)}{z q^2(1+\kappa G(0|0))},
\label{eq.lap.surv.pr}
\end{equation}
where 
\begin{align}
G(0|0) &= \frac{\ch(Lq)\ch(\alpha q)+\sh(Lq)\sh(\alpha q)z}{2qz(\ch(Lq)\sh(\alpha q)+\sh(Lq)\ch(\alpha q)z)} ,\nonumber \\
G(0|a) &= \frac{\ch(q(L+\alpha-a))}{2q(\ch(Lq)\sh(\alpha q)+\sh(Lq)\ch(\alpha q)z)} ,
\label{6}
\end{align}
$z=\exp(V_0)$, $L = \beta-\alpha$ and $q = \sqrt{s}$. We  first compute 
the survival probability $S_0(t)$ for the case where the width of the wells is 
large, i.e. $\lim L \rightarrow \infty$ limit. For this case 
the finite width of the 
barrier is of no consequence. So we assume that $\alpha \rightarrow 0$. 
The expression in \eq{eq.lap.surv.pr} becomes
\begin{equation}
\tilde{S}(s) = \frac{1}{q^2}-\frac{\kappa~\e^{-q a}}{z^2q^2(\kappa/z^2 + 2q)},
\label{lap.surv.pr.lim}
\end{equation}
The Inverse Laplace transform of \eq{lap.surv.pr.lim} gives the survival 
probability
\begin{equation}
S_0(t) = 1 - \mbox{erfc}\left(\frac{a}{2\sqrt{t}}\right) + \exp\left(\frac{a\tilde{\kappa}}{2}+\frac{\tilde{\kappa}^2t}{4}\right)\mbox{erfc}\left(\frac{a}{2\sqrt{t}}+\frac{\tilde{\kappa}\sqrt{t}}{2}\right),
\label{eq.surv.pr.lim}
\end{equation}
where $\tilde{\kappa}=\kappa/z^2$. This result can also be obtained 
from \eq{eq.lap.surv.pr} by substituting for the Green's function, 
$G(x|y) = \exp(-q|x-y|)/2q$ and setting the trapping rate 
$\kappa \rightarrow \kappa/z^2$. The result \eq{eq.surv.pr.lim} is  well 
known and it is the survival probability of a particle which is initially at 
position $x=a$ and diffuses to get trapped at the origin\cite{wi}.

The survival probability $S(t)$ for the general case can be obtained by 
computing the Inverse Laplace Transform of \eq{eq.lap.surv.pr}. The first 
term $1/q^2$ yields $1$. Substituting $q=\sqrt{s}$ in \eq{eq.lap.surv.pr}, 
the inverse Laplace transform of the second term can be written as
\begin{equation}
\frac{1}{2\pi i}\int^{\gamma+i\infty}_{\gamma-i\infty}\frac{G(0|a)\mbox{e}^{st}ds}{s(1+\kappa G(0|0))} = \frac{1}{2\pi i}\int^{\gamma+i\infty}_{\gamma-i\infty}\frac{\ch(\sqrt{s}(L+\alpha-a))\mbox{e}^{st}ds}{s (A(s)+\kappa'B(s))},
\label{con.intg}
\end{equation}
where $\kappa'=\kappa/z$, $A(s) = 2\sqrt{s}~[\ch(L\sqrt{s})\sh(\alpha \sqrt{s})+z\sh(L\sqrt{s}))\ch(\alpha \sqrt{s})]$,\\ 
$B(s) = \ch(L\sqrt{s})\ch(\alpha \sqrt{s})+z\sh(L\sqrt{s})\sh(\alpha \sqrt{s})$. The poles of the integrand in 
\eq{con.intg} are $s=0$ and the zeros of $A(s)+\kappa'B(s)$. The zeros $s^*$ satisfy 
\begin{equation}
\cos(L\rho)(\kappa'\cos(\alpha \rho)-2\rho\sin(\alpha \rho))-z\sin(L\rho)(\kappa'\sin(\alpha \rho)+2\rho\cos(\alpha \rho))=0,
\label{eq.root}
\end{equation}
where $\rho = \sqrt{|s^*|}$.

For the case $\kappa=0$, the denominator of the integrand in \eq{con.intg} 
becomes $sA(s)$. Although the integral for this case can be evaluated, one 
observes from  \eq{eq.lap.surv.pr} that its contribution to the survival 
probability shall be zero. Therefore, for $\kappa = 0$, $S(t) = 1$ for all 
$t>0$.

For $\kappa \neq 0$, the residue at $s=s^* \neq 0$ is $2\pi i \ch(\sqrt{s^*}(L
+ \alpha-a))\exp(s^*t)/[s^*\partial_s(A(s^*)+\kappa'B(s^*))]$. For $s=0$, 
the  residue is $2\pi i/\kappa'$. 
Using \eq{eq.lap.surv.pr} in \eq{con.intg} the expressions for the survival 
probability becomes
\begin{equation}
S(t)=\frac{\kappa}{z} \sum_\rho \frac{\cos(\rho(L+\alpha-a))}{F(\rho)}\mbox{e}^{-\rho^2 t},
\label{10}
\end{equation}
where $F(\rho)=s^*\partial_s(A(s^*)+\kappa'B(s^*))|_{s^*=-\rho^2}$. The values 
of $\rho$ can be computed using \eq{eq.root}.
\begin{figure}
\begin{center}
\includegraphics[width=0.5\textwidth]{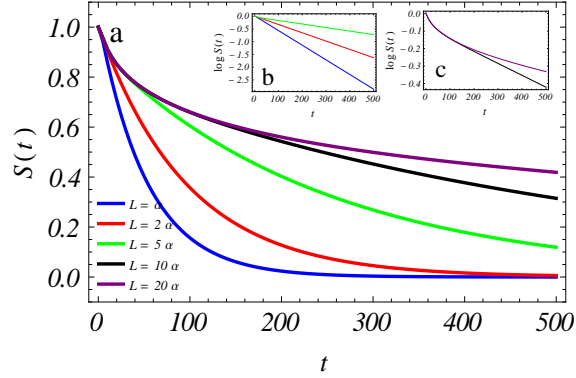}
\caption{Survival probability: (a) $S(t)$ decays with time for values of well 
width $L=\alpha,\ldots,10\alpha$, inset (b) for small $L=\alpha,2\alpha$ $\log S(t)$ 
is liner which implies an exponential decay and at large $L =5\alpha,10\alpha$ 
(c) it deviates from linearity (see the explanation in the text). 
Parameters used are $\alpha=2,~a=3,~V_0/k_BT=1,~\kappa =1$.}
\label{fig.surv.pr}
\end{center}
\end{figure}
\begin{figure}
\begin{center}
\includegraphics[width=0.5\textwidth]{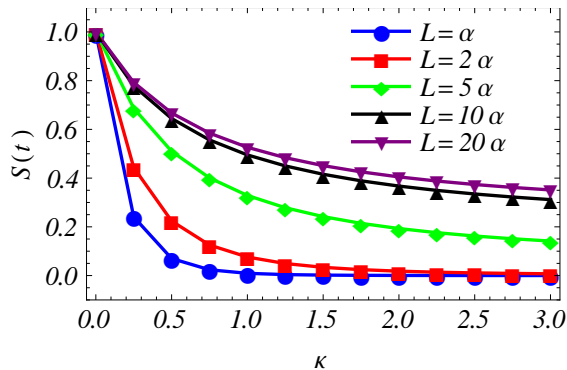}
\caption{Survival probability: $S(t)$ vs $\kappa$ at large time $t=250$ for 
values of well width $L=\alpha,\ldots,10\alpha$. Parameters used are 
$\alpha=2,~a=3,~V_0=1, \kappa =1$.}
\label{fig.surv.pr2}
\end{center}
\end{figure}

In \fig{fig.surv.pr}(a) we plot survival probability $S(t)$ as a function of 
time. For small values of the well width  (i.e. $L \sim \alpha$) we note that 
it decays exponentially (inset (b) where $\log S(t)$ has a linear behavior). 
However, for large well width $L \gg \alpha$ the survival probability 
$S(t)$ deviates from a single  exponential decay as seen in \fig{fig.surv.pr}. 
The explanation is as follows. With the range of  time, $t$ keeping fixed, 
when the length of the well is increased, contributions start coming 
not only from the lowest eigenvalue, but also from other near by 
eigenvalues. So, there will be a deviation from a single exponential 
behavior as seen in this calculation. This we have already 
seen in \eq{eq.surv.pr.lim} for the limiting case $L \rightarrow \infty$ 
(see inset \fig{fig.surv.pr}(c)). In the limit $L \rightarrow \infty$, 
eigenvalues form a quasi-continuum spectra and the 
limiting behavior is obtained by integrating over the spectral 
region. Furthermore, the survival probability at 
long time decays monotonically as a function of trapping rate $\kappa$ for 
all $L$ with the rate of decay smaller for larger $L$(see \fig{fig.surv.pr2}). 
This is due to the fact that a particle has the probability of moving 
away from a trap for a long time consequently reducing the over all decay 
probability.
\subsection{Short time behavior}
To calculate the survival probability at small time we need to expand the 
Green's function at large values of $q$ or $s$.
The Green's function for large $q$ can be written as
\begin{equation}
G(x|y) \simeq \frac{\mbox{e}^{-q|x-y|}}{q}\left\{\begin{array}{ll}
                   1/2z  &\mbox{~~~if~~~} |x|<\alpha,|y|<\alpha,\\
                   1/(1+z)&\mbox{~~~if~~~} |x|<\alpha,|y|>\alpha,\\
                   1/(1+z)&\mbox{~~~if~~~} |x|>\alpha,|y|<\alpha,\\
                   1/2 &\mbox{~~~if~~~} |x|>\alpha,|y|>\alpha,\\
                  \end{array} 
            \right.
\label{approx.green}
\end{equation}
Substituting in \eq{eq.lap.surv.pr} we obtain 
\begin{equation}
\tilde{S}(s) = \frac{1}{q^2}-\frac{\kappa~\e^{-q a}}{z(1+z)q^2(\kappa/2z + q)},
\label{approx.lap.surv.pr}
\end{equation}
The Inverse Laplace Transform yields
\begin{equation}
S(t) \simeq 1 - \frac{2}{1+z}\left\{\mbox{erfc}\left(\frac{a}{2\sqrt{t}}\right) - \exp\left(\frac{\kappa a}{2z}+\frac{\kappa^2 t}{4z^2}\right) \mbox{erfc}\left(\frac{a}{2\sqrt{t}}+\frac{\kappa\sqrt{t}}{2z}\right)\right\}.
\label{surv.pr.smallt}
\end{equation}
We note that the survival probability of a particle depends only on the 
scaled barrier height $V_0$, the trapping rate $\kappa$ and the distance 
from the the initial position to the trap $a$. The effect of confinement can 
be seen in the asymptotic time limit. Near $t=0$, $S(t)$ can be written as 
\begin{equation}
S(t) \simeq 1-\frac{4\kappa}{\sqrt{\pi}a^2(1+z)z}t^{3/2}\mbox{e}^{-a^2/4t}.
\end{equation}

\subsection{Long time behavior}
In the limit $L \rightarrow \infty$ and in the asymptotic time limit $t \rightarrow \infty$ the survival 
probability for the limiting case \eq{eq.surv.pr.lim} shows a power law decay. 
Expanding $S_0(t)$ for large values of $t$  we obtain
\begin{equation}
S_0(t)\sim\frac{1}{\sqrt{\pi}}\left(a + \frac{2}{\tilde{\kappa}}\right)t^{-1/2}.
\label{eq.asym.surv.pr}
\end{equation}
Let us now consider the general case where $L$ is finite.  Near  $s = 0$ 
we have 
\begin{align}
G(0|0) &\sim \frac{1-(L^2+\alpha^2-2zL\alpha) s/2)}{2sz[\alpha+Lz-\alpha L(\alpha z + L)s/2]} \nonumber \\
G(0|a) &\sim \frac{1+(L+\alpha-a)^2 s}{2s[\alpha+Lz-\alpha L(\alpha z + L)s/2]}.
\label{eq.asymp.lap.surv.pr}
\end{align}
Substituting \eq{eq.asymp.lap.surv.pr} expression in \eq{eq.lap.surv.pr} we 
obtain the asymptotic survival probability 
\begin{equation}
S(t) \simeq \left(1-\frac{\kappa'(L+\alpha-a)^2}{4(\alpha + Lz)+\kappa'(L^2+\alpha^2+2\alpha L z)}\right)\exp({\frac{-\kappa't}{2(\alpha + Lz)+\kappa'(L^2+\alpha^2+2\alpha L z)/2}}).
\label{eq.asymp.surv.pr}
\end{equation}
\section{Trapping reaction for a growing population}
\label{trap2}
For a vanishing growth term in \eq{1} particles eventually get trapped as they 
cannot escape the confining potential. Therefore, we cannot have a nonzero 
population surviving in the asymptotic large time limit. However, with a 
nonzero growth i.e. for a predator-prey system, the population may sustain 
itself in the asymptotically large time  regime. It would be interesting to 
find a threshold predation  rate $\kappa = \kappa_c$ above which the 
population may lead towards extinction.
 
\subsection{Trapping with linear growth}
Let us consider the case of linear growth
\begin{equation}
 \dt u = \dxs^2 u - \dxs(\phi u)-\kappa \delta(x) u + \lambda u.
\label{9}
\end{equation}
It can be shown that \eq{9} can be reduced to that of the trapping of a single 
particle with density $u'(x,t) = u(x,t)\exp(-\lambda t)$. As a result the 
total population $N(t) = \int u(x,t) dx$ can be written as the product of 
the survival probability $S(t)$ and $\exp(\lambda t)$. 

In the limiting case we have $N(t) = S_0(t)\exp(\lambda t)$. Using \eq{eq.asym.
surv.pr} the asymptotic population can be written as
\begin{equation}
\lim_{t\rightarrow \infty} N(t) =: N_{\infty}(t) \simeq \frac{1}{\sqrt{\pi}}\left(a + \frac{2}{\tilde{\kappa}}\right)t^{-1/2}\exp(\lambda t).
\label{10}
\end{equation}

Clearly, the population diverges for all $\tilde{\kappa}>0$. Hence, there 
exist no threshold $\kappa_c$ for a linear growth model in the limiting 
case $L \rightarrow \infty$. 
In other words, for a large habitat size, localized predation cannot 
drive a linearly growing population extinct.

\begin{figure}
\begin{center}
\includegraphics[width=0.5\textwidth]{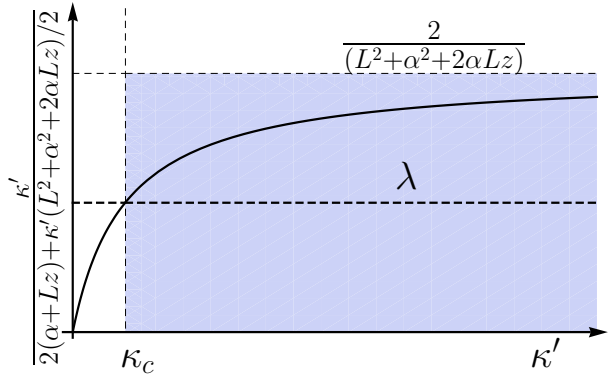}
\caption{Extinction of population: Intersection of the curve with the dashed 
horizontal line at point $(\kappa_c, \lambda)$ correspond to exponent zero 
in \eq{eq.asym.popl.finiteL}. In the shaded region $\kappa'>\kappa_c$ and 
$0<\lambda<2/(L^2+\alpha^2+2 \alpha L z)$ the population becomes extinct.}
\label{fig.plot.extinct}
\end{center}
\end{figure}

Similarly, for $L$ finite using \eq{eq.asymp.surv.pr} we have 
\begin{equation}
N_{\infty}(t) \sim \exp({\frac{-\kappa't}{2(\alpha + Lz)+\kappa'(L^2+\alpha^2+2\alpha L z)/2}}+\lambda t).
\label{eq.asym.popl.finiteL}
\end{equation}
We observe from \eq{eq.asym.popl.finiteL} that there exist a threshold rate 
$\kappa' = \kappa_c$ above which the population becomes extinct(see \fig{fig.plot.extinct}).
Note that if $\lambda>2/(L^2+\alpha^2+2 \alpha L z)$ the population does not 
go extinct for all $\kappa'>0$. Note again the followings. (1) When $L\ 
\rightarrow\ infty$ the population  diverges irrespective the value of 
$\kappa$  for any positive value of $\lambda$. Similarly, for 
$V_{0}\rightarrow \infty$, the population diverges for positive values of 
$\lambda$. Here the reason is that in the limit $V_{0}\ \rightarrow 
\infty$, habitats are confined to their respective wells.

The threshold trapping rate can be written as 
\begin{equation}
\kappa_c = \frac{2(\alpha + Lz)\lambda}{1-(L^2+\alpha^2+2 \alpha L z)\lambda/2}
\label{eq.thres.kappa}
\end{equation}
This is an important result of this work. Note that when $L \rightarrow 
\infty$, $\kappa_{c}\ \rightarrow\ 0.$
\subsection{Trapping with logistic growth}
We shall compute the population $N(t)$ for a logistic growth $F(u) = \lambda 
u(1-u/\bar{u})$ where $\bar{u}$ is the carrying capacity. However, unlike the 
linear growth case here we cannot compute population exactly due to the 
nonlinearity. We obtain a perturbative solution for finite time and at steady 
state.

\subsubsection{Perturbative solution at finite time:}
Inserting the expression for $F(u)$ in \eq{1} we have
\begin{equation}
 \dt u = D\dxs^2 u - \dxs(\phi u)-\kappa \delta(x) u + \lambda (1-u/\bar{u})u.
\label{trap.log0}
\end{equation}
Now dividing \eq{trap.log0} through by $\bar{u}$ and using the transformation
$t\rightarrow Dt$, $\kappa \rightarrow \kappa/D$, $\lambda \rightarrow \lambda/
D$ and $u \rightarrow u/\bar{u}$ we can write
\begin{equation}
 \dt u = \dxs^2 u - \dxs(\phi u)-\kappa \delta(x) u + \lambda (1-u)u,
\label{trap.log}
\end{equation}
with the boundary condition $\lim_{|x|\rightarrow \infty} u =0$ and initial 
condition $u(x,0)=\delta(x-a)/\bar{u}$. Let $u = u_0 + \lambda u_1 + \ldots$ 
so that we have 
\begin{align}
\mathcal{O}(1):~~~ &  \dt u_0 = \dxs^2 u_0 - \dxs(\phi u_0)-\kappa \delta(x) u_0\\
\mathcal{O}(\lambda):~~~ &  \dt u_1 = \dxs^2 u_1 - \dxs(\phi u_1)-\kappa \delta(x) u_1 + (1-u_0)u_0,
\end{align}
with initial conditions $u_0(x,0)=\delta(x-a)/\bar{u}$ and $u_1(x,0)=0$. 
Note that u$_{1}$ has the dimension of (length)$^{2}$.
Taking the Laplace Transform and by using the Green's function one can 
write the solution $\tilde{u}_0$, the Laplace Transform of u$_{0}$  as 
(see \eq{3})
\begin{equation}
\tilde{u}_0 = \frac{1}{\bar{u}}\left(G(x|a) - \frac{\kappa G(x|0)G(a|0)}{1+\kappa G(0|0)}\right).
\label{sol.ord0}
\end{equation}
Note that $\tilde{u}_0$ has the dimension of (length)$^{2}$ as required.
Similarly, we can write 
\begin{equation}
\tilde{u}_1 = \frac{-\kappa \tilde{R}(0,s)G(0|x)}{1+\kappa G(0|0)} + \tilde{R}(x,s),
\label{sol.ord1}
\end{equation}
where $\tilde{R}(x,s) = \int \tilde{r}(x',s) G(x'|x)dx'$ and $\tilde{r}(x,s)$ 
is the Laplace transform of $u_0(1-u_0)$. We note again that 
$\tilde{u}_1$ has the requisite dimension of (length)$_{2}$. Taking the 
Inverse Laplace Transform  of \eq{sol.ord0} gives the solution for single 
particle trapping case 
(see \eq{3}). 
The contribution due to the growth term up to   $\mathcal{O}(\lambda)$ can 
be computed from \eq{sol.ord1}. For large $q$ we can write $\tilde{u}_0$ as
\begin{equation}
\tilde{u}_0 \simeq \frac{1}{\bar{u}}\left\{\begin{array}{ll}
\frac{\mbox{e}^{-q|x-a|}}{q(1+z)}- \frac{\kappa~\exp(-q(|x|+a))}{2qz(1+z)(q+\kappa /2z)} \mbox{~~if~~} |x|<\alpha\\
\frac{\mbox{e}^{-q|x-a|}}{2q}- \frac{\kappa~\exp(-q(|x|+a))}{2q(1+z)(q+\kappa /2z)} \mbox{~~if~~} \alpha<|x|<\beta.
                     \end{array}
              \right.
\end{equation}
This is an approximate solution obtained by using \eq{approx.green} for the 
Green's function. Similarly the integrand $\tilde{r}(x,s)$ for large $q$ can 
be written as
\begin{equation}
\tilde{r}(x,s)\simeq \tilde{u}_0 - \frac{2}{\pi\bar{u}^2}K_0(\sqrt{2}|x-a|q)\left\{\begin{array}{ll}
                       \frac{1}{(1+z)^2}, \mbox{~~if~~} |x|<\alpha, \\
                       \frac{1}{4}, \mbox{~~~~~if~~} \alpha<|x|<\beta,
                     \end{array}
              \right.
\end{equation}
where $K_0(\cdot)$ is the modified Bessel function of the second kind. Although
further approximations can be made to compute the solution  
$u(x,t) \simeq u_0(x,t)+\lambda u_1(x,t)$ , the expression will be too 
complicated and will not be very useful. Therefore, instead of examining the 
behavior of $u$, we 
investigate that  of  the total population $N(t) = \bar{u}\int u dx$. 
The following integration gives the approximate total population 
\begin{align}
\tilde{N}(s) &= \bar{u}\int_{-\beta}^{\beta} u_0 dx + \lambda \bar{u}\int_{-\beta}^{\beta} u_1 dx + \mathcal{O}(\lambda\kappa), \nonumber \\
&\simeq \tilde{S}(s) + \lambda\bar{u}\int_{-\beta}^{\beta} \tilde{R}(x,s)dx,\nonumber \\
&\simeq \tilde{S}(s)\left(1+\frac{\lambda}{q^2}\right)-\frac{\lambda}{2\pi\bar{u}}\int_{-\beta}^{\beta}\int^{\beta}_{\alpha}K_0(\sqrt{2}|x'-a|q)G(x'|x)dx'dx.
\label{popl.log}
\end{align}
where $\tilde{S}(s)$ is defined by \eq{approx.lap.surv.pr}. 
In \eq{popl.log} only the dominant terms are retained as other terms are 
exponentially small. Integrating the Green's function we have 
$\int G(x'|x)dx \sim q^{-2}$. The integrand $K_0(\sqrt{2}|x'-a|q) \sim -(\gamma
+\log(q|x'-a|/\sqrt{2}))$ if $a-c/q < x < a+c/q$ for all $q$ where $\gamma = 0.
577216\ldots$ is the Euler constant and $c = \sqrt{2}\exp(-\gamma)$. For $|x'-a
|>c/q$ the approximation for $\tilde{r}_1(x',t)$ is negative. The integral 
becomes
\begin{equation}
\int_{-\beta}^{\beta}\int^{\beta}_{\alpha}K_0(\sqrt{2}|x'-a|q)G(x'|x)dx'dx = \int_{-\beta}^{\beta}\int^{a-c/q}_{a+c/q}K_0(\sqrt{2}|x'-a|q)G(x'|x)dx'dx \simeq \frac{2c}{q^3}.
\label{33}
\end{equation}
We note from \eq{popl.log} that the first term corresponds approximately to 
that of the linear growth as in \eq{10} and (\ref{eq.asym.popl.finiteL}) and 
from \eq{33} the contribution $\lambda c/(\pi \bar{u} q^3)$ reduces the 
population by an amount $\sim 2\lambda c\sqrt{t}/(\pi^{3/2} \bar{u})$.

\subsubsection{Steady state solution at low trapping and growth rates:}
\begin{figure}
\centering
\includegraphics[width=0.5\textwidth]{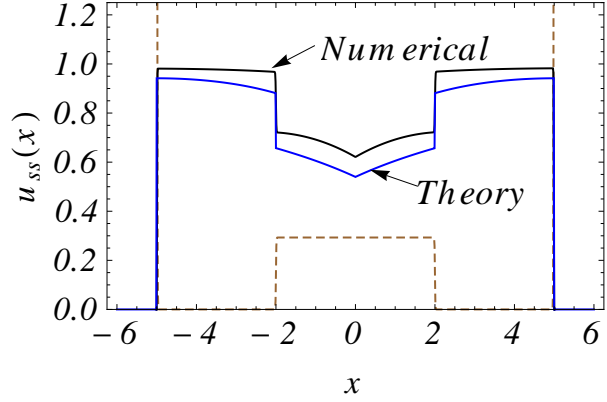}
\caption{Steady state solution $u_{ss} \simeq u_0+\kappa u_1$ (blue curve)compa
red with the numerical solution for parameter $\lambda = 0.2$, $\kappa = 0.2$ 
and $V_0 = 0.3$.The potential(brown dashed line) is described by the 
parameters $\alpha = 2$, $\beta=5$ and initial condition $u(x,0)$ (black curve)
 is chosen as a delta function at  $x=3$ which is then evolved up to time 
$t=40$ where solution converge to a steady state solution (black curve).}
\label{f2}
\end{figure}

The solution to the steady state equation
\begin{equation}
\dxs^2 u - \dxs(\phi u)-\kappa \delta(x) u + \lambda u(1-u)=0,
\label{diffeq.trap.growth}
\end{equation}
gives the density in the asymptotic long time limit. So, its 
solution can 
be used to determine the population $N_\infty$. First, we consider the case 
where both $\lambda, \mbox{and} \kappa \ll 1$.  
Let $u_0$ be the solution to \eq{diffeq.trap.growth} for $\kappa = 0$, $\lambda
 = 0$. Using the boundary conditions $\dxs u(\pm \beta) = 0$ and jump condition
s\cite{morsch} $\dxs u(x^-) = \dxs u(x^+)$, $\exp(V(x^-))u(x^-) = \exp(V(x^+))u
(x^+)$ where $x=\pm \alpha$ we obtain
\begin{equation}
u_0(x) = \left\{\begin{array}{ll}
                   1  &\mbox{~~~if~~~} \alpha<|x|<\beta,\\
                   1/z &\mbox{~~~if~~~} |x|<\alpha.
                  \end{array} 
            \right.
\end{equation}
Now, let us write $u = u_0 + \kappa u_1 + \kappa^2 u_2 + \ldots$ which on 
substitution into \eq{diffeq.trap.growth} gives
\begin{equation}
\mathcal{O}(\kappa) :~~ \dxs^2 u_1 - \dxs(\phi u_1) - \lambda (2u_0-1)u_1 = \delta(x) u_0.
\label{13}
\end{equation}
We note that \eq{13} has solution of the form $a \exp(k x)+ b \exp(-k x)$ with 
$k=\sqrt{\lambda(2u_0-1)}$. Furthermore, we have $\lambda(2u_0-1) = \lambda$ 
if $\alpha<|x|<\beta$ and $\lambda(2/z-1)$ if $|x|<\alpha$. The solution $u_1$ 
can be written as
\begin{equation}
u_1(x) = \frac{-1}{2\Delta(0)}\left\{\begin{array}{ll}
                   \ch(k_1(\beta-|x|))  &\mbox{~~~if~~~} \alpha<|x|<\beta,\\
                   \Delta(|x|)/qz &\mbox{~~~if~~~} |x|<\alpha.
                  \end{array} 
            \right.
\end{equation}
where $\Delta(x) = k_2 \ch(k_2(\alpha-|x|))\ch(k_1L)+k_1z\sh(k_2(\alpha-|x|))\sh(k_1L)$, $k_1 = \sqrt{\lambda}$ and $k_2 = k_1\sqrt{2/z-1}$.
A comparison of the approximate steady state solution $u_{ss} \simeq u_0+\kappa u_1$ with the numerical solution is shown in \fig{f2}.
Integration of  the steady state solution gives the asymptotic population
\begin{equation}
N_\infty = 2\bar{u}(L+\alpha/z) - \frac{\kappa \bar{u}}{\Delta(0)}\left( \sh(k_1L)\left(\frac{1}{k_1}-\frac{1}{k_2}\right) + \frac{\sh(k_1\alpha)\ch(k_1L)}{zk_1}+ \frac{\ch(k_1\alpha)\sh(k_1L)}{k_2}\right).
\label{n.inf}
\end{equation}
Using the smallness of the parameters \eq{n.inf} becomes
\begin{equation}
N_\infty \simeq \bar{u}\left(L+\alpha \text{e}^{-V_0}\right)\left(2- \frac{\kappa}{\sqrt{\lambda}}\right).
\end{equation}
We note that depletion due to localized predation is proportional to the ratio 
$\kappa/\sqrt{\lambda}$.
\section{Numerical results}
\label{num}
We consider a quartic double well potential described by $V(x) = V_0(1-x^2/a^2)^2$. 
The potential is symmetric and has a maximum at the origin and  minima 
at $x=\pm a$. The central barrier is of height $V_0$. We choose this smooth 
double well potential to numerically investigate the model for arbitrary 
values of parameter $\lambda$ and $\kappa$. The choice of this smooth 
$\phi^{4}$ potential also stems from our interest 
  to examine the agreement of results, at least qualitatively, obtained 
from this potential  with results, obtained from the square well 
bistable potential. Furthermore, this $\phi^{4}$ potential is widely used standard for any bistable system. 

\begin{figure}
\centering
\includegraphics[width=0.7\textwidth]{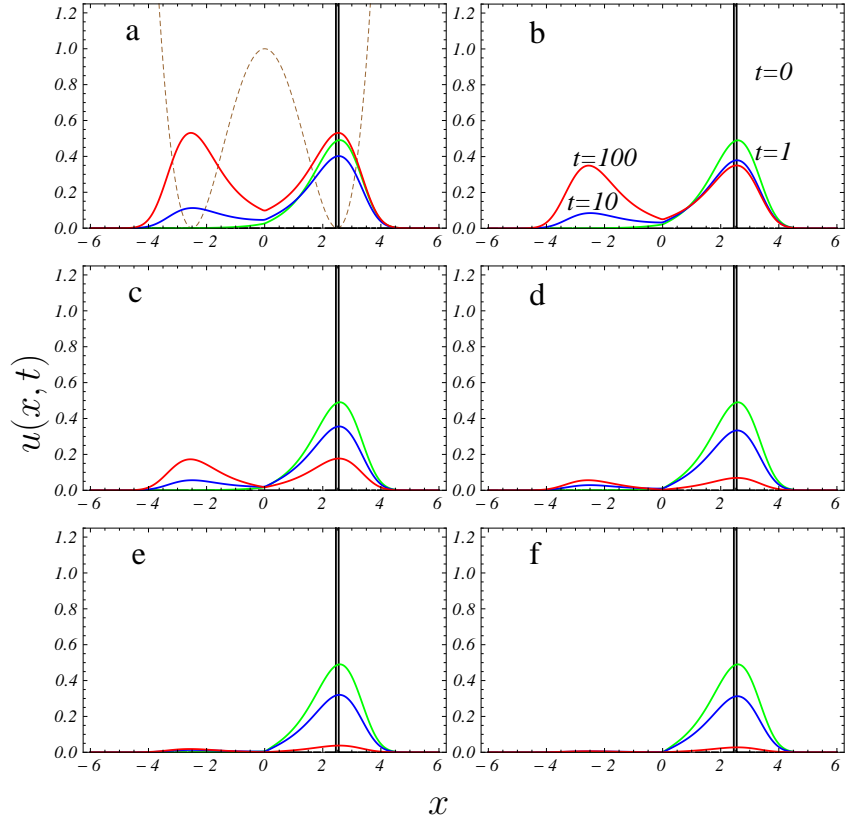}
\caption{Evolution of density $u(x,t)$ as a function of time and space. The 
dashed curve in (a) shows the double well potential with parameters $V_0=1$, 
$a=2.5$. The growth rate $\lambda = 0.1$ and the trapping  rates are (a) 
$\kappa = 1.0$, (b) $\kappa = 1.5$, (c)$\kappa=2.5$, (d)$\kappa = 5.0$, 
(e)$\kappa=10.0$ and (f)$\kappa=20.0$.}
\label{density1}
\end{figure}

\begin{figure}
\centering
\includegraphics[width=0.5\textwidth]{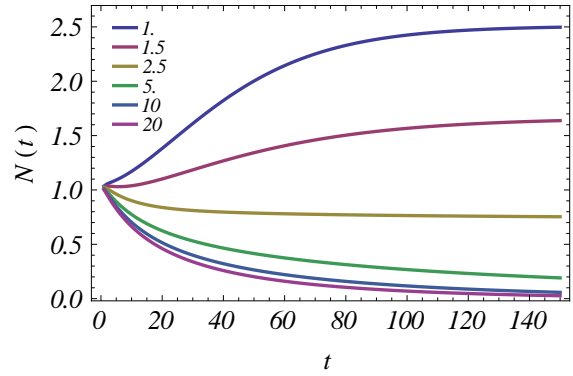}
\caption{Total population as a function of time for $\kappa = 1.0,\ldots,20$. 
Parameters used are $\lambda = 0.1$, $V_0 = 1.0$, $a=2.5$ and initial 
population is $N(0) = 1$.}.
\label{tot.popl}
\end{figure}

We consider a particle initially located in the right well at $x=a$. 
In \fig{density1} we have plotted the evolution of 
the density $u(x,t)$ at time $t=0,1,10$ and $100$ for various values of 
trapping rate $\kappa$. We observe that at large $t$ the density approaches 
a steady state density, as we have seen our bistable square well system.  
The steady state density is nonzero for small trapping 
rate and approaches zero as we increase the trapping rate. Again it is 
in agreement with our analytical results for our bistable square well 
system.  We can see this in \fig{tot.popl} where 
total population $N(t)$ at large time tends to zero for large values of 
trapping rate and approaches a finite value for sufficiently small values. 
This behavior can be explained by the equation,
\begin{equation} 
dN/dt =  -\kappa_{\text{eff}} N  + \lambda' N(1-N),
\end{equation}
where $\lambda' = \lambda/\bar{u}$ is the growth rate.
This can be obtained from \eq{diffeq.trap.growth} if we 
assume that the effects of diffusion and trapping on the total 
population can be clubbed together by replacing the total effect by an 
effective decay rate $\kappa_{\text{eff}}$. The 
solution $N(t)=(1-\kappa_{\text{eff}}/\lambda')[1-\kappa_{\text{eff}}\exp(\kappa_{\text{eff}}t-\lambda' t)/\lambda']^{-1}$ has the same behavior 
as \fig{tot.popl}. The effective decay rate can be written in terms of 
the asymptotic population $N_\infty$ as $\kappa_{\text{eff}} = \lambda'(1-N_\infty)$.

\begin{figure}
\centering
\includegraphics[width=0.5\textwidth]{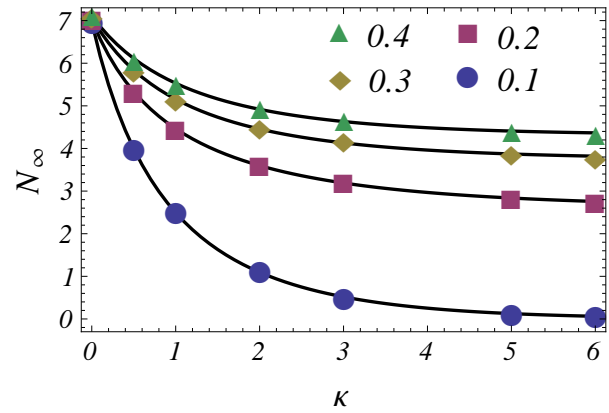}
\caption{Population in the asymptotic time limit as a function of $\kappa$ for 
$\lambda=0.1,0.2,0.3$ and $0.4$ with $V_0=1$, $a=2.5$.}
\label{asymp}
\end{figure}

The asymptotic population for various values of $\lambda$ and $\kappa$ is 
shown in \fig{asymp}. By least square fitting we found that 
$N_\infty(\kappa) = p/(q + \mbox{e}^{r\kappa})$ where $p, q$ and $r$ are constants. 
The parameter $r$ is positive for small values of $\lambda$ 
where $N_\infty \rightarrow 0$ for large $\kappa$. However, as 
 $\lambda$ increases, $r$ changes continuously from positive to negative 
values. So, when $\kappa \gg 1/|r|$, $ N_{\infty}\ \rightarrow p/q$.  
Note that then $\kappa_{\text{eff}}$ can 
go from a positive to a negative value. This, in turn , leads to a 
nonzero steady state population. Similar threshold behavior were 
predicted analytically for the square double well potential case.   
\section{Conclusion}
In this paper we studied the trapping reaction problem in a symmetric double 
well potential. A trap located at the middle of the central barrier of the 
double well potential is considered. This is done in the context of ecology 
where the confining potential is modelled as the habitat and the trap as 
localized predation. We observed that due to the confinement the asymptotic 
survival probability decay exponentially which in the absence of the 
potential shows a power law behavior. Furthermore, trapping reaction of a 
linearly growing population was studied where it is shown that even for 
an arbitrily large 
predation (trapping) the population does not vanish at long time. However, 
in presence of a confining potential, for a given range of growth rate there 
exist a threshold trapping rate $\kappa_c$ above which the population becomes 
extinct in the asymptotic time limit. For a logistic growth term we computed a 
first order perturbative solution for the total population. We also found from 
the steady state solution  that asymptotic population depletes by a fraction 
$\kappa/\sqrt{\lambda}$. 
Numerical studies are done for the case of a quartic potential an the results 
are discussed.

There can be many variations of this model. First of all, instead of a box 
potential, one can consider a finite height potential. Potential can have 
various interesting shapes. Furthermore, it is possible to have more than 
one routes to connect the habitats. In this case, the time evolution of 
survival probability can show interesting behavior. We consider all these 
problems in our subsequent analysis.   

\appendix
\section{Derivation of Green's function}
\label{appndx}
The Green's function $G(x|y)$ is defined by
\begin{equation}
(s- \dxs^2 + \dxs \phi)G(x|y) = \delta(x-y),
\end{equation}
with boundary condition $G(x|y) = 0$ as $x \rightarrow \infty$. In abstract 
notation we can write it as $G = [s- \dxs^2 + \dxs \phi]^{-1}$.
For the double square well potential $\phi=0$ except at points $x=\pm \alpha$ 
and $\pm \beta$. Therefore the Greens function take the form
\begin{equation}
G = A \mbox{e}^{q x} + B \mbox{e}^{-q x},
\end{equation}
where $A$ and $B$ are constant that depends on $y$ and $q=\sqrt{s}$. At the 
point $\tilde{x}=\alpha$ the Green's function satisfy the jump conditions 
$G(\tilde{x}^+) = \exp(V_0)G(\tilde{x}^-)$, $\dxs G(\tilde{x}^+) 
= \dxs G(\tilde{x}^-)$ and at point $\tilde{x}=\beta$ we have
 $\dxs G(\tilde{x}|y) =0$\cite{morsch}. Similarly, jump conditions for the 
points $x=-\alpha$ and $-\beta$ are imposed. These jump conditions along 
with the continuity conditions at point $x=y$ gives

\begin{equation}
G(x|y) = \frac{1}{q\Delta}\left\{\begin{array}{ll}
                   g(x|y) &\mbox{~~~if~~~} x<y\\
                   g(y|x) &\mbox{~~~if~~~} x>y,
                  \end{array} 
            \right.
\label{79}
\end{equation}
where 
\begin{align}
g(x|y) = \left\{\begin{array}{ll}
                    \ch(q(\beta+x))F_0(y), 
 &\mbox{~if~}  -\beta<x<y<-\alpha,\\
                    \ch(q(\beta+x))F_3(y), &\mbox{~if~} -\beta<x<-\alpha<y<\alpha,\\
                    F_3(y) F_3(-x), &\mbox{~if~} -\alpha<x<y<\alpha,\\
                    \ch(q(\beta-y))F_3(-x), &\mbox{~if~} -\alpha<x<\alpha<y<\beta,\\
                    \ch(q(\beta-y))F_0(x), &\mbox{~if~~~}\alpha < x < y < \beta,\\
                    \ch(q(\beta-y))\ch(q(x+\beta))z, &\mbox{~if~} -\beta<x<-\alpha<\alpha<y<\beta,
                  \end{array} 
            \right.
\label{89}
\end{align}
with $\Delta = \sinh(2q\alpha)(\cosh^2(qL)+\sinh^2(qL)z^2) + \cosh(2q\alpha) \
sinh(2qL)z$,$~L=\beta-\alpha$,
\begin{align}
F_0(x)&=\sh(2q\alpha)F_1(x)+\ch(2q\alpha)F_2(x)z,\\
F_1(x)&=\sh(q(|x|-\alpha))\ch(qL)+\ch(q(|x|-\alpha))\sh(qL)z^2,\\
F_2(x)&=\ch(q(2\alpha-\beta-|x|)),\\
F_3(x)&=\ch(q(x+\alpha))\ch(qL)+\sh(q(x+\alpha))\sh(qL)z.
\end{align}

\end{document}